\documentclass[10pt,letterpapers,floats]{article}
\usepackage{amsfonts,color,epsfig}
\usepackage{multirow}
\usepackage[multiple]{footmisc}
\usepackage{booktabs}
\newcommand{\nc}{\newcommand}

\newcommand{\rnc}{\renewcommand}
\renewcommand{\thefootnote}{\fnsymbol{footnote}}
\makeatletter
\rnc{\theequation}{\thesection.\arabic{equation}}
\@addtoreset{equation}{section}
\makeatother
\nc{\fig}[5]{ 
\begin{figure}[!htbp]
    \begin{center}
    \leavevmode
    \centerline{
        \includegraphics[width=#1, height=#2]{#3}
        }
    \caption[]{#4}
    \label{#5}
    \end{center}
\end{figure}}
\nc{\figs}[8]{
\begin{figure}[!htbp]
    \begin{center}
    \leavevmode
    \centerline{
        \includegraphics[width=#1, height=#2]{#3}
        \includegraphics[width=#4, height=#5]{#6}
         }
    \caption[]{#7}
    \label{#8}
    \end{center}
\end{figure}}

\begin{document}
\begin{flushright}
\end{flushright}
\vspace{14mm}
\begin{center}
{{{\Large {\bf Formation of Five-Dimensional String Solutions from the Gravitational Collapse}}}}\\[10mm]
{Seungjoon Hyun$^{a,c}$\footnote{email:hyun@phya.yonsei.ac.kr},
Jaehoon Jeong$^{a}$\footnote{email:jejechi@gmail.com},
Wontae Kim$^{b,c}$\footnote{email:wtkim@sogang.ac.kr}, and 
John J. Oh$^{a}$\footnote{email:john5@yonsei.ac.kr}}\\[10mm]

{{${}^{a}$ Institute of Physics and Applied Physics, Yonsei University, Seoul, 120-749, Korea\\[0pt]       
${}^{b}$ Department of Physics, Sogang University, Seoul, 121-742, Korea\\[0pt]
${}^{c}$ Center for Quantum Space-time, Sogang University, Seoul, 121-742, Korea}\\[0pt]
}
\end{center}
\vspace{4mm}
\begin{abstract}
We study the formation of five-dimensional string solutions including the Gregory-Laflamme (GL) black string, the Kaluza-Klein (KK) bubble, and the geometry with a naked singularity from the gravitational collapse. The interior solutions of five-dimensional Einstein equations describe collapsing
non-isotropic matter clouds. It is shown that the matter cloud always forms the
GL black string solution while the KK bubble solution cannot be formed. 
The numerical study seems to suggest that the collapsing matter forms the geometries with timelike naked curvature singularities, which should be taken cautiously as the general relativity is not reliable in the strong curvature regime.  

\end{abstract}
\vspace{5mm}



\vspace{1.3cm}

\hspace{11.5cm}{Typeset Using \LaTeX}
\newpage
\renewcommand{\thefootnote}{\arabic{footnote}}
\setcounter{footnote}{0}
\section{Introduction}\label{sec:intro}
The idea of spacetimes of dimensions more than four has been around for more than eighty years since the Kaluza-Klein's suggestion. It has been successfully adopted by string theory and brane-cosmology to date. It is also interesting to study physics with extra dimensions in the context of general theory of relativity as well.
In general relativity in higher dimensions, various black hole solutions with a spherical symmetry are found and studied in detail. However, apart from these solutions, there is an interesting class of black objects, namely, black string solutions with a translational symmetry along an extra coordinate. 
The simplest case found by Gregory and Laflamme \cite{gl} is a black string solution in five-dimensions which is a direct product of four-dimensional Schwarzschild black hole with extra circle. It has been shown that the solution has an unstable mode under small perturbations \cite{gl2} and this solution can evolve to a new type of nonuniform solutions \cite{hm} from some numerical studies. See some reviews of ref. \cite{ho1,ho2,kol,hno} on this issue and references therein.

Motivated by this, a new class of string solutions was
recently rediscovered with two independent parameters; energy density
and tension density along the fifth direction \cite{chl}. In
particular, it includes the Gregory-Laflamme (GL) black string
solution and the Kaluza-Klein (KK) bubble solution. These solutions of
the five-dimensional Einstein equation were firstly discovered in
ref. \cite{gp} in the viewpoint of KK magnetic monopoles. Afterwards,
these were rediscovered in studying four-dimensional black holes in
the presence of one extra dimension \cite{do}. Besides these, these
solutions were investigated mostly in the context of KK theory in some
literatures \cite{kra,yos,bro}. In addition, a new stationary solution
with a conserved momentum along the fifth direction was recently found \cite{lk}.

However, geometrical properties such as causal structures of this generic class of string solutions have not been well-understood yet despite of their peculiar and intriguing characters. 
In this sense, it would be very interesting to study whether or not a certain configuration of the matter distribution can form these solutions from the gravitational collapse in the context of general relativity.

In this paper, we study the formation of these solutions from the gravitational collapse for generic parameters of tension density. In section \ref{sec:bstension}, some properties of these string solutions are discussed. In section \ref{sec:interior}, the matching of the interior space of the cloud of matters in the five-dimensional cylindrical coordinate system and the string solutions with arbitrary tensions as an exterior space outside the cloud are considered. 
A matching condition on the edge along the fifth direction leads to a 
restriction of the interior metric ansatz and consistent equations of motion.
The interior solutions on the edge describing a collapsing non-isotropic matter cloud within finite time are studied in a numerical way.
In section \ref{sec:match}, the matching conditions between extrinsic curvatures giving the dynamics of the cloud edge are considered and equations of motion describing the dynamics of the cloud edge are obtained.
In section \ref{sec:sols}, it is shown that the equations of motion can be solved numerically by setting some specific values of parameters. The qualitative behaviors of the cloud edge show that the matter cloud can collapse to the GL black string solution, depending on the parameters associated with the tension along the fifth direction. However we find that the KK bubble solution cannot be a final state of gravitational collapse for the matter distributions we considered. We also discuss the gravitational collapse toward the geometry with a naked singularity.  
Finally, we summarize and discuss our results in section \ref{sec:discussion}.

\section{Some Properties of String Solutions with Arbitrary Tension}\label{sec:bstension}
We consider a class of string solutions with an arbitrary tension in five dimensions, which has a metric in the isotropic coordinates as \cite{chl}
\begin{equation}
\label{eq:metricbs}
(ds)^2 = -F(\rho)dT^2 + G(\rho)(d\rho^2 + \rho^2 d\Omega^2)+H(\rho)dz^2, 
\end{equation}
where
\begin{eqnarray}
&&F(\rho) = \left(1-\frac{K_\omega}{\rho}\right)^{s}\left(1+\frac{K_\omega}{\rho}\right)^{-s},\\
&&G(\rho) = \left(1-\frac{K_\omega}{\rho}\right)^{2-\frac{1+\omega}{2-\omega}s}\left(1+\frac{K_\omega}{\rho}\right)^{2+\frac{1+\omega}{2-\omega}s},\\
&&H(\rho) = \left(1-\frac{K_\omega}{\rho}\right)^{-\frac{1-2\omega}{2-\omega}s}\left(1+\frac{K_\omega}{\rho}\right)^{\frac{1-2\omega}{2-\omega}s},
\end{eqnarray}
with
\begin{equation}
s=\frac{2(2-\omega)}{\sqrt{3(\omega^2-\omega+1)}},~~K_\omega = \sqrt{\frac{\omega^2-\omega+1}{3}}G_5 M.
\end{equation}
A constant $\omega$ represents an equation-of-state parameter defined as the ratio of the string tension density $\tilde{\tau}$ to the mass density $M$,
\begin{equation}
\tilde{\tau} \equiv -\int dx^3 T_{44} = \omega \int dx^3 T_{00} \equiv \omega M,
\end{equation}
and $G_5$ is a five-dimensional Newton's constant. The range of $\omega$ is given by $0\le \omega \le 2$ to guarantee a robust physical ground of the strong energy condition (SEC) ($\omega\le 2$) defined by
\begin{equation}
T_{\mu\nu}\xi^{\mu}\xi^{\nu} + \frac{1}{3} T^{\mu}_{~\mu} \ge 0,
\end{equation}
where $\xi\cdot\xi <0$ as well as the positivity of tension, $\tilde{\tau}$ ($\omega\ge 0$) \cite{tras}. For $\omega=2$ ($\alpha=0$), the metric (\ref{eq:metricbs}) describes the KK bubble solution \cite{kkbub} while $\omega=1/2$ ($\alpha=1$) describes the GL black string solution \cite{gl}. Note that $\rho$ spans from zero to infinity for $\omega=2$ while it spans from $K_{\omega}$ to infinity for other values of $\omega$. 

In terms of radial coordinate $R=\rho\sqrt{G}$, the metric (\ref{eq:metricbs})  can be expressed in an alternate form of
\begin{equation}
\label{eq:radial}
(ds)^2 = - FdT^2 + \left(1+\frac{\rho}{2G}\frac{dG}{d\rho} \right)^{-2} dR^2 + R^2 d\Omega^2 + H dz^2.
\end{equation}

The Ricci scalar curvature vanishes since the metric is a vacuum solution of the five-dimensional Einstein equation. However, the curvature singularity determined by the Kretschmann scalar is located at $\rho=K_{\omega}$ in all range of $\omega$, except for $\omega=1/2, 2$ while it is clearly finite in the whole range of $\rho$ for $\omega=1/2,2$. 
In the case of $\omega =1/2$, the curvature singularity exists at $R=0$, which is not covered by the isotropic coordinate, $\rho$.

An event horizon is defined by a surface where an outgoing light toward the radial direction is trapped,
\begin{equation}
\frac{d\rho}{dT} = \rho^2\frac{(\rho-K_{\omega})^{\alpha}}{(\rho+K_{\omega})^{\alpha+2}} = 0,
\end{equation}
with $\alpha\equiv \sqrt{3}/\sqrt{\omega^2-\omega+1}-1$.
This seems to suggest that an apparent horizon exists at $\rho=K_{\omega}$ for any $0\le \omega <2$ ($\alpha>0$). On the other hand, there is a naked singularity at $\rho=K_{\omega}$ for other values of $\omega$, which will be shown later. For $\omega=2$ describing the KK bubble solution, the geometry is geodesically complete in the range of $\rho
\ge K_{\omega}$. The range of $0\le \rho \le K_{\omega}$ is identical to that of $\rho\ge K_{\omega}$. In what follows, we consider the whole range of $0\le \rho \le \infty$ for conveniences. Some properties of these solutions are classified in the Table \ref{table1}.

\begin{table*}[htdp]
\caption{Some properties of black string solutions for varying $\alpha$ (or $\omega$) in the isotropic coordinate system}
\begin{center}
\begin{tabular}{|c|c|c|c|c|c|}
\hline
~~solutions~~ & ~~$\alpha$~~ &~~ $\omega$~~ & ~range of $\rho$~ & apparent horizon & curvature singularity \\
\hline\hline
 KK bubble solution & $0$ & $2$ & $K_{\omega}\le \rho < \infty$ &$\cdot$ & $\cdot$  \\  \hline
 GL black string~ & $1$ & $1/2$ & $K_{\omega}\le \rho < \infty$ &$\rho=K_\omega$ & $R=0$\\ \hline
 Sols. with NS & $0<\alpha<1$ & others & $K_{\omega} \le \rho < \infty$& $\rho=K_{\omega}$ & $\rho=K_\omega$   \\  \hline
\end{tabular}
\end{center}
\label{table1}
\end{table*}

\section{Matching Conditions and Collapsing Clouds Inside}\label{sec:interior}
Let us first assume the matter clouds in five-dimensional spacetime manifolds, then the regions are separated into two regions of inside and outside clouds, denoted by ${\mathcal V}_{-}$ and ${\mathcal V}_{+}$, respectively. Each regions are described by the interior metric $g_{\mu\nu}^{-}$ and the coordinate system $\{t,r,\theta, \phi, z\}$ inside and the exterior metric $g_{\mu\nu}^{+}$ and the coordinate system $\{T, R, \theta, \phi, z\}$ outside, respectively. In order to investigate a collapsing cloud inside, one needs to study matching conditions on the hypersurface and the metric of the interior spacetimes. The matching conditions which should be imposed in our case are
\begin{equation}
[g_{ij}]=0,~~[K_{ij}]=0,\label{eq:matchcond}
\end{equation}
where $[A] \equiv A_{{\mathcal V}_+}-A_{{\mathcal V}_-}$ and $K_{ij}$ is an extrinsic curvature to the tangential directions of the matter cloud defined by
\begin{equation}
K_{ij} = - n_{\sigma}(\partial_{j}e^{\sigma}_{(i)} + \Gamma_{\mu\nu}^{\sigma}e^{\mu}_{(i)}e^{\nu}_{(j)}),
\end{equation}
where $n_{\sigma}$ is a normal vector and $e^{\sigma}_{(i)}$ is a basis vector to the hypersurface.

Taking into account the matching condition for the ($zz$)-component of the extrinsic curvature, $[K_{zz}]=K_{zz}^{+} - K_{zz}^{-} = 0$, gives
\begin{equation}
\frac{\sqrt{F}\dot{T}}{2\sqrt{G}}\frac{\partial_{\rho}H}{H} = \frac{n_{r}^{-}}{2}g^{rr}_{-}\partial_{r}g_{zz}^{-} (e^{z}_{(z)})^2,\label{eq:zzexcur}
\end{equation}
where the overdot ($\cdot$) represents $d/dt$.

Provided $g_{zz}^{-}=g_{zz}^{-}(t)$, then the rhs in eq. (\ref{eq:zzexcur}) vanishes and thus we have $\dot{T}=0$. It implies that there is no radial component of the normal vector to the hypersurface, which results in the breakdown or inconsistency of equations of motion. We, therefore, wish that the lhs in eq. (\ref{eq:zzexcur}) does not vanish ($\dot{T}\ne 0$) and should impose $g_{zz}^{-} = g_{zz}^{-}(t,r)$. This implies that the given static geometry (\ref{eq:metricbs}) cannot be obtained by the gravitational collapse of a homogeneous matter distribution, assuming the Friedmann-Robertson-Walker (FRW) metric inside. This is a minimal restriction of taking a metric ansatz for the inside geometry.  
Therefore, we take our metric ansatz inside the cloud in the form of
\begin{equation}
\label{eq:inmet}
(ds)^2 = -dt^2 + a^2(t)(dr^2+r^2d\Omega^2)+ \xi^2(t,r)dz^2
\end{equation}
where $a(t)$ is a scale factor and $\xi^2(t,r) = h^2(t) b^2(r)$. 

The five-dimensional Einstein equation, $G_{\mu\nu}=\kappa^2 T_{\mu\nu}$,
where $T_{~\nu}^{\mu} = -diag(\lambda,\tau,\tau,\tau,\tau_5)
$ and $\kappa^2 = 8\pi G_5$, yields
\begin{eqnarray}
&&\frac{1}{ra^2 h b}\left[ 3r\dot{a}^2 hb - rhb'' + 3rba\dot{h}\dot{a} - 2hb'\right] = \kappa^2 \lambda,\label{eq:tt}\\
&&\frac{1}{ra^2hb}\left[2rahb\ddot{a} + rhb\dot{a}^2 + 2rab\dot{h}\dot{a} + ra^2b\ddot{h} - 2hb'\right] = \kappa^2\tau, \label{eq:rr}\\
&& \frac{1}{ra^2hb}\left[2rahb\ddot{a} + rhb\dot{a}^2 + 2rab\dot{h}\dot{a} + ra^2b\ddot{h} - hb'-rhb''\right] = \kappa^2 \tau, \label{eq:thethe}\\
&& \frac{3}{a^2}\left[\dot{a}^2 + a\ddot{a}\right] = \kappa^2 \tau_5,\label{eq:55}\\
&& \frac{b'}{hba}\left[a\dot{h}-\dot{a}h\right] = 0, \label{eq:tr}
\end{eqnarray}
where the overdot ($\cdot$) and the prime ($\prime$) denote $d/dt$ and $d/dr$, respectively.
Note that $\lambda$ is an energy density, $\tau$ ($\tau_5$) is a tension along $i$-direction ($z$-direction), where $i=r,\theta,\phi$.
From eq. (\ref{eq:tr}), $h(t)=a(t)$ can be chosen since $b'(r)\ne 0$ is assumed.
Subtracting eq. (\ref{eq:thethe}) from eq. (\ref{eq:rr}) gives $rb''-b'=0$, which has a general solution,
\begin{equation}
b(r)=r_1 r^2 + r_2,
\end{equation}
where $r_1$ and $r_2$ are integration constants. For simplicity, we take $r_2=0$, hereafter. Then, the equations of motion are simplified to
\begin{eqnarray}
&& \frac{\dot{a}^2}{a^2} - \frac{1}{a^2 r^2} = \frac{\kappa^2}{6}\lambda(t,r),~~\frac{\dot{a}^2}{a^2}+\frac{\ddot{a}}{a} = \frac{\kappa^2}{3} \tau_5(t), \label{eq:EQ12}\\
&& \tau(t,r) = \tau_5 (t) - \frac{4}{\kappa^2 r^2 a^2}. \label{eq:EQ3}
\end{eqnarray}

On the other hand, the energy-momentum conservation, $\nabla_{\mu}T^{\mu\nu}=0$, yields two equations,
\begin{equation}
\dot{\lambda} + \frac{\dot{a}}{a}(4\lambda - 3\tau - \tau_5) = 0,~~~\tau' + \frac{2}{r}(\tau-\tau_5) =0.
\end{equation}
Note that the first equation is a continuity equation with respect to the energy flow while the second one is equivalent to eq. (\ref{eq:EQ3}). It is hard to obtain an analytic solution of above equations because of the deficiency of the number of equations for given variables. 
More precisely, for given six variables, there are only three independent equations. Therefore, the additional gauge choice of fixing $r=r_0$ on the edge of the cloud is required in order to solve the equations of motion.
Indeed, it is sufficient to investigate the qualitative motion of the cloud edge governed by the scale factor in the interior spacetimes. On the edge of the cloud at $r=r_0$, equations of motion, eqs. (\ref{eq:EQ12}) and (\ref{eq:EQ3}), are rewritten as
\begin{eqnarray}
&&\frac{\dot{a}^2}{a^2} - \frac{1}{r_0^2 a^2} = \frac{\kappa^2}{6}\lambda_e,~~
\frac{\ddot{a}}{a} +\frac{1}{r_0^2a^2} = \frac{\kappa^2}{6}(2\tau_5-\lambda_e),\label{eq:EQS12}\\
&&\tau_e = \tau_5 - \frac{4}{\kappa^2r_0^2a^2},\label{eq:EQS3}
\end{eqnarray}
where $\lambda_e\equiv\lambda(t,r_0)$ and $\tau_e\equiv\tau(t,r_0)$.

Provided we assume $\tau_5 = c \lambda_e$, where $c$ is a constant, then the continuity equation becomes 
\begin{equation}
\label{eq:conteq}
\dot{\lambda_e} + \frac{\dot{a}}{a} \left[4(1-c)\lambda_e + \frac{3k_0}{a^2}\right] = 0,
\end{equation}
where $k_0 = 4/\kappa^2 r_0^2$. In addition, subtracting the first equation from the second equation in eq. (\ref{eq:EQS12}) yields 
\begin{equation}
\label{eq:eqs}
\frac{d}{dt}\left(\frac{\dot{a}}{a}\right) = \frac{\kappa^2}{3}(c-1)\lambda_e -\frac{2}{r_0^2a^2}.
\end{equation}
At $t=t_0=0$, eq. (\ref{eq:eqs}) gives a positivity condition of an initial energy density, $\lambda_{0}^{e}$, by
\begin{equation}
\label{eq:posden}
\lambda_{0}^{e} = \frac{3}{\kappa^2(c-1)} \left[\frac{r_0^2(a_0\ddot{a}_0 - \dot{a}_0^2)+2}{r_0^2a_0^2}\right] > 0,
\end{equation}
which produces a condition $r_0^2(a_0\ddot{a}_0 - \dot{a}_0^2) +2< 0$ for the matter satisfying the null energy condition (NEC) along the fifth-direction, $\lambda_e -\tau_5 \ge 0$ (i.e. $c \le1$).
Defining $X \equiv \ln(a/a_0)$ and plugging eq. (\ref{eq:eqs}) into (\ref{eq:conteq}), one finds
\begin{equation}
\label{eq:deqn}
\frac{d}{dt}\ddot{X} - 4(c-1)\ddot{X}\dot{X} - \frac{4c}{a_0^2r_0^2} \dot{X}e^{-2X} = 0.
\end{equation}
It is hard to obtain an exact solution of eq. (\ref{eq:deqn}) in an analytic manner due to its severe nonlinearity and complexity. Instead, numerical solutions of the scale factor can be found by a set of initial conditions and the positivity condition of the energy density. 

At $t=t_0=0$, $X_0 = 0$ since $a(t=0)=a_0$ while $\dot{X}_0 = \dot{a}_0/a_0$ will be chosen to be negative to have the collapsing behavior initially. For example, if $a_0=1$ and $\dot{a}_{0}=-1$, then we have $\dot{X}_0 = -1$. Finally, $\ddot{X}_{0} = (a_0\ddot{a}_0-\dot{a}_{0}^2)/a_0^2$ can be chosen from an initial acceleration, $\ddot{a}_0$, satisfying the positivity condition of the energy density in eq. (\ref{eq:posden}) (for example, $\ddot{a}_0 = -2$ for $r_0=1$ and $c=0.5$). For these initial data, the differential equation can be numerically solved.\footnote{The fourth-order Runge-Kutta method was used here.}

On the other hand, the Ricci scalar curvature at $r=r_0$ defined by
\begin{equation}
{\mathsf R}_{~\mu}^{\mu}[\Sigma] = \frac{4}{a^2r_0^2}\left(2r_0^2 a\ddot{a} + 3r_0^2\dot{a}^2 - 3\right)
\end{equation}
diverges as the scale factor approaches to zero.
Some numerical plots for the scale factor, the energy density, and the Ricci scalar curvature are  illustrated in Fig. \ref{fig:numscale2}. 
 
In the case of $c = 1$, an analytic solution of the scale factor can be found to be
\begin{equation}
a(t) = a_0 \left[2 e^{-\frac{\sqrt{a_0\ddot{a}_0-\dot{a}_0^2}}{\sqrt{2}a_0}t}-e^{\frac{\sqrt{a_0\ddot{a}_0-\dot{a}_0^2}}{\sqrt{2}a_0}t}\right].
\end{equation} 
In this case, provided $\ddot{a}_0 a_0 - \dot{a}_0^2 > 0$, then the solution clearly collapses to a point. The behavior of the solution is similar to the plots in Fig. \ref{fig:numscale2}.

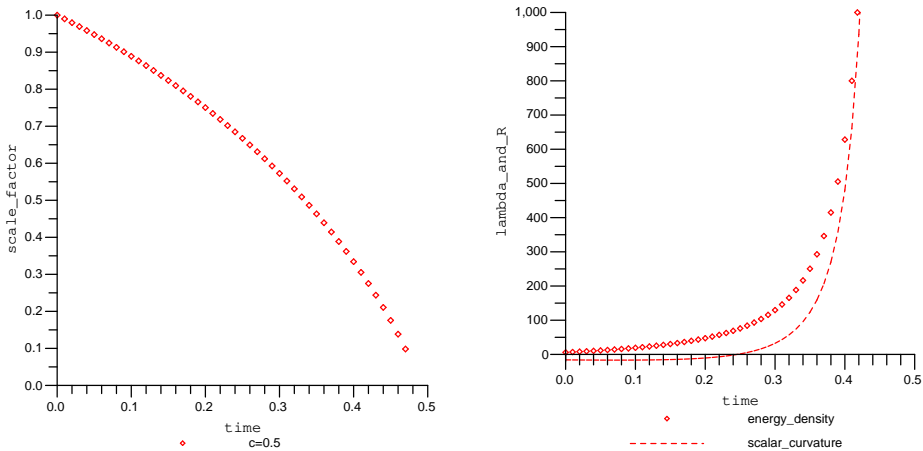
\figs{6.5cm}{6.5cm}{scalefactors}{6.5cm}{6.5cm}{energyden}{\small Numerical behaviors of the scale factor (lhs) and the energy density and the scalar curvature (rhs) of the collapsing matter cloud edge when $a_0=1$, $\dot{a}_0=-1$, $\ddot{a}_0=-2$, and $r_0=1$.}{fig:numscale2}

\section{Matching on the Edge and the Equations of Motion}\label{sec:match}

From the matching condition for metrics, eq. (\ref{eq:matchcond}), the exterior and the interior metrics on the edge of the matter cloud at $\rho=\varrho(t)$ and $r=r_0$, respectively, can be written in the form of
\begin{eqnarray}
(ds)^2_{\Sigma} &=& -F(\varrho)dT^2 + G(\varrho)(d\varrho^2 + \varrho^2 d\Omega^2)+H(\varrho)dz^2 \label{eq:metedge}\\
&=& - dt^2 + r_0^2 a^2(t)d\Omega^2 + a^2(t)b^2(r)dz^2,
\end{eqnarray}
which yields 
\begin{equation}
\label{eq:matches}
F\dot{T}^2 = G\dot{\varrho}^2+1,~~ G\varrho^2 = r_0^2a^2,~~H = a^2b^2.
\end{equation}
The extrinsic curvature tensor is defined by
\begin{equation}
K_{ab} = - n_{\alpha}\frac{\delta e^{\alpha}_{(a)}}{\delta \sigma^{b}}=-n_{\alpha}(\partial_{b}e_{(a)}^{\alpha} + \Gamma_{\mu\nu}^{\alpha}e_{(a)}^{\mu}e_{(b)}^{\nu}),
\end{equation}
where $n_{\mu}$ is the normal to the edge of the clouds, $e_{(a)}^{\alpha}$ are the basis vectors on the edge, and $\sigma^{\alpha}$ are the coordinates on the edge. In the interior coordinates, the basis vectors are 
\begin{eqnarray}
&&e_{(t)}^{\mu} = (1,0,0,0,0),~~e_{(\theta)}^{\mu} = \left(0,0,\frac{1}{r_0 a(t)},0,0\right),\nonumber\\
&&e_{(\phi)}^{\mu} = \left(0,0,0,\frac{1}{r_0a(t)\sin\theta},0\right),~~e_{(z)}^{\mu} = \left(0,0,0,0,\frac{1}{a(t)b(r)}\right)
\end{eqnarray}
and the unit normal vector, $n_{\mu}=(0,a(t),0,0,0)$ while in the exterior coordinates, they are
\begin{eqnarray}
&&e_{(T)}^{\mu} = (\dot{T},\dot{\varrho},0,0,0),~~e_{(\theta)}^{\mu} = \left(0,0,\frac{1}{\varrho\sqrt{G}},0,0\right),\nonumber\\
&&e_{(\phi)}^{\mu} = \left(0,0,0,\frac{1}{\varrho\sqrt{G}\sin\theta},0\right),~~e_{(z)}^{\mu} = \left(0,0,0,0,\frac{1}{\sqrt{H}}\right)
\end{eqnarray}
and the unit normal vector is $n_{\mu} = (-\sqrt{GF}\dot{\varrho},\sqrt{GF}\dot{T},0,0,0)$.

The non-vanishing components of the extrinsic curvatures in the interior spacetimes are evaluated to be
\begin{equation}
K_{\theta\theta}^{-} = K_{\phi\phi}^{-} = \frac{1}{r_0 a},~~K_{zz}^{-} = \frac{2}{r_0 a},
\end{equation}
while they are also found in the exterior coordinates,
\begin{eqnarray}
&&K_{tt}^{+} = -\frac{1}{\sqrt{GF}}\partial_{\varrho} \sqrt{GF\dot{\varrho}^2+F},\\
&&K_{\theta\theta}^{+}=K_{\phi\phi}^{+} = \frac{\sqrt{F}\dot{T}}{\varrho\sqrt{G}}\left(1+\frac{\varrho}{2G}\partial_{\varrho}G\right),\\
&&K_{zz}^{+} = \frac{\sqrt{F}\dot{T}}{2\sqrt{G}} \frac{\partial_{\varrho}H}{H}
\end{eqnarray}

From the second matching condition for extrinsic curvatures $[K_{ij}]=0$ in eq. (\ref{eq:matchcond}), one finds
\begin{eqnarray}
&&\frac{d}{d\varrho} \sqrt{GF\dot{\varrho}^2+F} = 0, \label{eq:rhoeqs}\\
&&\frac{\sqrt{F}\dot{T}}{\varrho\sqrt{G}}\left(1+\frac{\varrho}{2G}\frac{dG}{d\varrho}\right)-\frac{1}{r_0a}=0,~~\frac{\sqrt{F}\dot{T}}{2\sqrt{G}H} \frac{dH}{d\varrho}-\frac{2}{r_0a}=0 \label{eq:rel2}.
\end{eqnarray}

The dynamics of the cloud edge in time is determined by solving eq. (\ref{eq:rhoeqs}), which can be rewritten in terms of the effective potential,
\begin{equation}
\label{eq:rhode}
\dot{\varrho}^2 + V_{eff}(\varrho) = 0,
\end{equation}
with
\begin{equation}
V_{eff}(\varrho) = \frac{F(\varrho)-F_0 (1+G_0V_0^2)}{G(\varrho)F(\varrho)},
\end{equation}
where $G_0\equiv G(\varrho_0)$, $F_0\equiv F(\varrho_0)$, and $V_0\equiv \dot{\varrho}_0 = \dot{\varrho}(0)$. Two equations in eq. (\ref{eq:rel2}) give functional relations between metric functions inside and outside the edge. More precisely, it is found that both equations of eq. (\ref{eq:rel2}) are identical using eq. (\ref{eq:matches}). Since we have ${F}\dot{T}=const.$ from eq. (\ref{eq:rhoeqs}) and the solutions for $a(t)$ and $\varrho(t)$ are related to $c$ and $\omega$, respectively, eq. (\ref{eq:rel2}) 
gives some relations between parameters $c$ and $\omega$. However, since we do not find the exact solutions in an analytic way, it is not easy to show this relation explicitly.

The effective potential diverges at $\varrho=K_{\omega}$ excluding the case of $\omega=2$. The shape of the effective potential and possible collapse scenarios are shown in Fig. \ref{fig:veff}. 
For these generic values of $0\le \omega <2$, the edge will either approach to $\varrho=K_{\omega}$ or expand indefinitely (the dash-dot line in Fig. \ref{fig:veff}), depending on its initial condition. Provided the effective potential has one root (the dash line in Fig. \ref{fig:veff}), then the edge will collapse to a point at $\varrho=K_\omega$ all the time. In the collapsing case, the time-evolution of the edge will be terminated at $\rho=K_{\omega}$.

For $\omega=2$ (the solid line in Fig. \ref{fig:veff}), the effective potential always has a negative value except at the origin, which implies that the collapsing edge will approach to the origin since there is no horizon. 
The KK bubble solution is properly described in the radial coordinate system. The radial coordinate $R$ covers the whole geometry and spans from the minimal radius at $R=R_{min}=4K_{\omega}$, which corresponds to the position at $\rho=K_{\omega}$ in the isotropic coordinate system. In addition, the asymptotic behaviors at $\rho=0$ and $\rho=\infty$ in the isotropic coordinate is identical to that of $R=\infty$ in the radial coordinate. Therefore the isotropic coordinate system doubly covers the KK bubble geometry and $\rho=0$ and $\rho=\infty$ describes the same asymptotic region. This implies that the KK bubble geometry can never be a final state for the matter distribution we considered. Even if the edge starts contracting, it will expand indefinitely after contracting to the minimal radius at $R=R_{min}$.

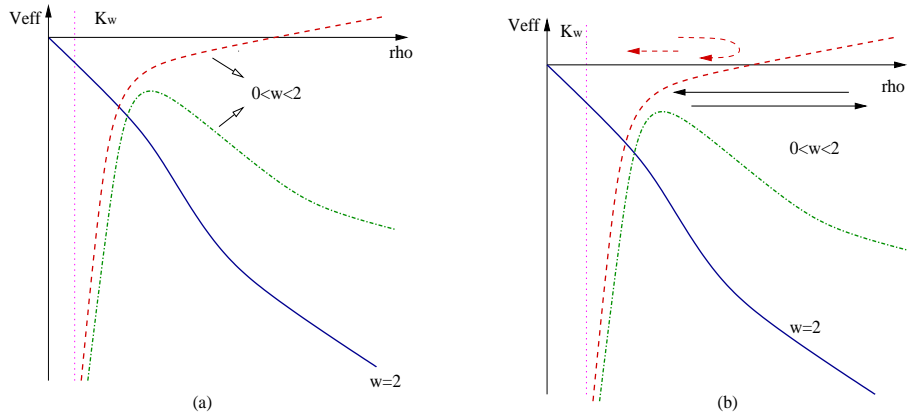
\fig{12cm}{5.5cm}{veff}{Plots of the effective potential for values of $\omega$ (a) and possible collapse scenarios for each case (b), depending on its initial condition.}{fig:veff}

\section{Formation of String Solutions: Numerical Results}\label{sec:sols}

As seen above, it is too difficult to obtain analytic solutions of eq. (\ref{eq:rhode}) for an arbitrary parameter, $\omega$, because of its severe nonlinearity. However, some numerical solutions can be found by setting appropriate initial conditions, which is sufficient to figure out the qualitative behavior of solutions.
The differential equation, (\ref{eq:rhode}), is solved by setting some initial conditions, $\varrho_0=2$, $V_0^2\equiv\dot{\varrho}_0^2 = 1$, $G_5=1$, and $M=1$. Figs. \ref{fig:rhosol1}, \ref{fig:rhosol2}, and \ref{fig:rhosol3} show the behaviors of the cloud edge in time for some different value of $\omega$. As seen in all figures, the edge of the cloud shrinks to a point at $\varrho=K_{\omega}$, except for $\omega=2$. 

\subsection{The GL Black String Case: $\alpha=1$}
Since the coordinate spans from the event horizon at $\rho=K_{\omega}$ to infinity in the case of $\alpha=1$ ($\omega=1/2$) describing the GL black string, the edge of the cloud will evolve until it meets the horizon within finite time and the evolution will be terminated. In the isotropic coordinate we worked, one cannot see the formation of the singularity inside the horizon. However, this can be achieved through the similar analysis in the radial coordinate system. Then, the edge will form an event horizon within finite time, keep shrinking, and form a curvature singularity at $R=0$ screened by an event horizon. More precisely, the metric in the radial coordinate system, (\ref{eq:radial}) on the edge at $R={\mathcal R}(t)$ becomes
\begin{equation}
\label{eq:radialedge}
(ds)^2_{\rm edge} = - FdT^2 + \left(1+\frac{\rho}{2G}\frac{dG}{d\rho} \right)^{-2} d{\mathcal R}^2 + {\mathcal R}^2 d\Omega^2 + H dz^2.
\end{equation}
Then, the velocity of the edge, $\dot{\mathcal R}$, in the radial coordinate is given by
\begin{equation}
\label{eq:vele}
\dot{\mathcal R}^2 = - V_{eff}(\varrho({\mathcal R})) \left(1-\frac{K_{\omega}^2}{\varrho^2({\mathcal R})}\right)^2,
\end{equation}
which is clearly finite at the horizon. In addition, since the acceleration of the edge is $\ddot{{\mathcal R}} = - 2K_{\omega}\varrho^2({\mathcal R})/(\varrho+K_{\omega})^4$, the edge will keep evolving into the horizon.
This will give a usual result of the black string formation from the gravitational collapse of matter, which is depicted for some initial data of $\varrho_0=2$, $V_0=-1$, $M=1$, and $G_{(5)}=1$ in the lhs of Fig. \ref{fig:rhosol1}. See the appendix \ref{sec:appendix} on the formation of the GL black string from the collapsing matter in the radial coordinate.

\subsection{The KK Bubble Case: $\alpha=0$}
For $\omega=2$ (the KK bubble solution), the rhs of Fig. \ref{fig:rhosol1} shows that the edge approaches to a point at $\varrho=0$ in infinite time. Since, as alluded earlier, the KK bubble solution has a minimal radius, $R_{min}=4K_{\omega}$, that is equivalent to $\rho=K_{\omega}$, the edge can shrink to the point at $R=R_{min}$ ($\rho=K_{\omega}$). At this point, the velocity of the edge vanishes while the effective potential has a negative values in all range of $R>R_{min}$ in the radial coordinate due to eq. (\ref{eq:vele}).

Therefore the point at $\rho=K_{\omega}$ ($R=R_{min}=4K_{\omega}$) is a metastable position, at which the edge will either stay or bounce back toward infinity, depending on its acceleration. The acceleration of the edge is given by
\begin{equation}
\ddot{\mathcal R} = \frac{2K_{\omega}V_0^2 (\varrho_0+K_{\omega})^4}{\varrho_0^4} \frac{\varrho^2}{(\varrho+K_{\omega})^4} > 0,
\end{equation}
which has an maximum at $\varrho=K_{\omega}$, at which the acceleration starts to decrease. Therefore, we see that the edge once reached at $\rho=K_{\omega}$ will bounce back toward infinity at $\rho=0$ ($R=\infty$), which implies that the KK bubble solution will not be a final state of the gravitational collapse in our configurations.

\subsection{The Generic String Solution Case: $0<\alpha<1$}

For all values except for $\omega=1/2,2$, the edge always collapses to a point at $\rho=K_{\omega}$ and forms a curvature singularity at that point within finite time.\footnote{This behavior is similar to the typical property of an attractor in that the time-evolution of a function always approaches to a certain fixed point, irrespective of any initial parameters.} To figure out whether or not the curvature singularity is time-like, one needs to evaluate the coordinate time observed by an external observer, yielding
\begin{equation}
T=T_0 + \int_{\rho_{0}}^{\rho_{h}} \frac{(\rho+K_\omega)^{\alpha+2}}{\rho^2(\rho-K_\omega)^{\alpha}}d\rho.
\end{equation}
The power of $(\rho-K_\omega)$ in the denominator has a range of $0< \alpha \le 1$ for $0\le \omega < 2$: the maximum, $\alpha=1$, describes the GL black string for $\omega=1/2$ while $\alpha=0$ describes the KK bubble for $\omega=2$. The coordinate time diverges when $\alpha=1$ ($\omega=1/2$) since 
\begin{equation}
T=T_0 +\int_{\rho_{0}}^{\rho_{h}}  \frac{(\rho+K_\omega)^{3}}{\rho^2(\rho-K_\omega)}d\rho \sim 8K_{\omega}\ln(\rho-K_\omega)|_{\rho_0}^{\rho_h} \rightarrow \infty.
\end{equation}
This tells us that the light emitted from the horizon cannot approach to an observer's eyesight for the GL black string solution and the curvature singularity at origin is clearly cloaked by the event horizon. 

However, some numerical searches of the integration show that the coordinate times are clearly finite except for the case of $\omega=1/2$ ($\alpha=1$).
For example, when $\alpha=1/2$, the integration can be evaluated in an analytic form, but it is not singular at $\rho=K_{\omega}$. The mathematical reason on this is because the power of the denominator in the integrand is non-integer with $0<\alpha<1$. Physically, no event horizon is formed at $\rho=K_{\omega}$ despite of the existence of an apparent horizon at that point since the coordinate time at which an external observer detects the light emitted from an event horizon is finite. This implies that an external observer will see a curvature singularity at $\rho=K_\omega$ within finite time, which seems to suggest an emergence of a naked singularity. However, the numerical results based on the general relativity near the naked singularity, where the curvature diverges, can not be trusted as quantum effects presumably dominates in that region. 
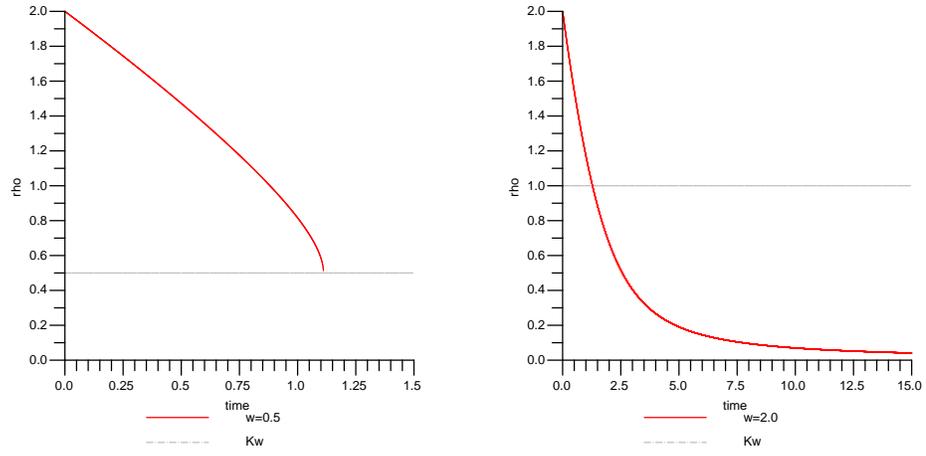
\figs{6.5cm}{6.5cm}{rho05}{6.5cm}{6.5cm}{rho20}{\small Numerical solutions of the cloud edge, $\varrho(t)$, for $\omega=1/2$ and $\omega=2$ when $\varrho_0=2$. The edge of the cloud approaches to $K_\omega$ for $\omega=1/2$ while it approaches to $\varrho=0$ (${\mathcal R}=\infty$) for $\omega=2$. The lhs is the GL black string and the rhs is KK bubble solution.}{fig:rhosol1}
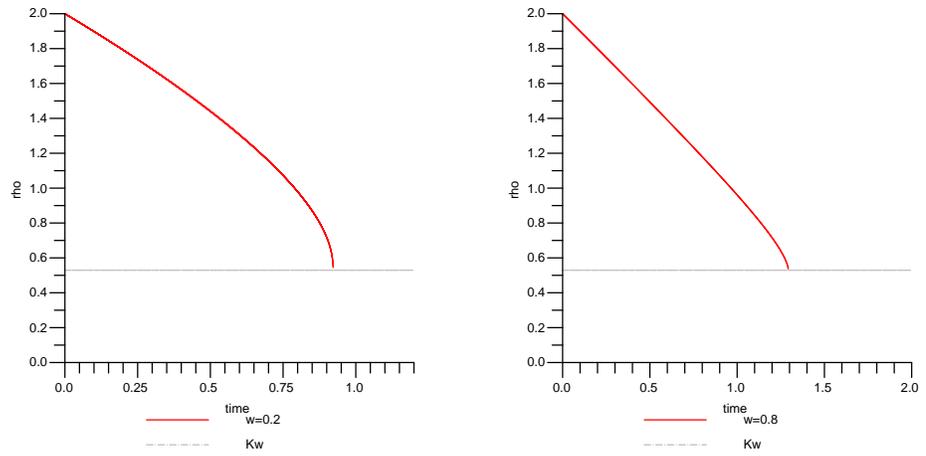
\figs{6.5cm}{6.5cm}{rho02}{6.5cm}{6.5cm}{rho08}{\small Numerical solutions of the cloud edge, $\varrho(t)$, for $\omega=0.2$ and $\omega=0.8$ when $\varrho_0=2$. The edge always approaches to $K_{\omega}$, regardless of its initial conditions.}{fig:rhosol2}
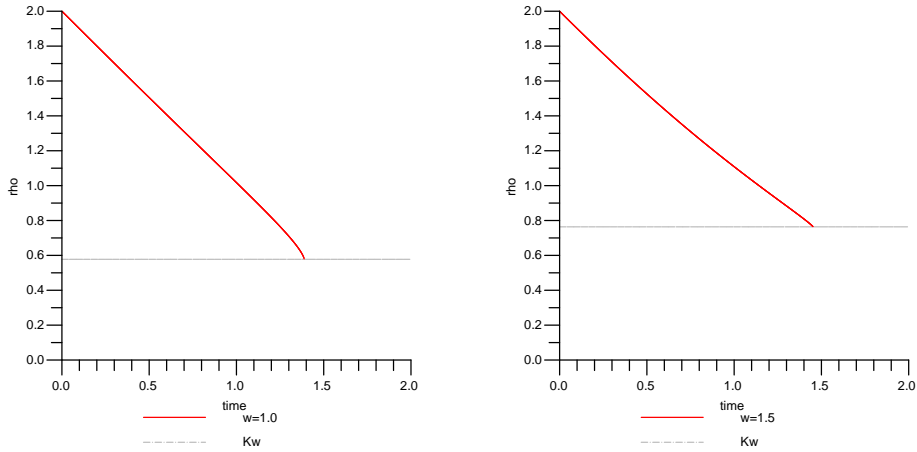
\figs{6.5cm}{6.5cm}{rho10}{6.5cm}{6.5cm}{rho15}{\small  Numerical solutions of the cloud edge, $\varrho(t)$, for $\omega=1.0$ and $\omega=1.5$ when $\varrho_0=2$. The edge always approaches to $K_{\omega}$, regardless of its initial conditions.}{fig:rhosol3}

\section{Discussion}\label{sec:discussion}

The main interest of this paper is to investigate the formation of string solutions with arbitrary tensions from the gravitationally collapsing non-isotropic cloud of matter. 
For generic values of $\omega$, the edge of the cloud collapses to a fixed point at $\rho=K_\omega$, which looks like a typical behavior of an attractor in the context of the chaotic nonlinear dynamics. However, their physical interpretations are very different. For the well-known solutions of the GL black string ($\omega=1/2$), the edge of the cloud approaches to their end point ($\rho=K_{\omega}$) and the time-evolution is terminated at that point without seeing the formation of a curvature singularity.
But the whole analysis should be done in the radial coordinate system of eq. (\ref{eq:radial}) for further time-evolutions. This can be easily achieved since their solutions are known in the radial coordinate system. As a result, we found that the edge of the cloud forms an event horizon and then a curvature singularity at $R=0$ for the GL black string.

More interesting case is for the KK bubble solution. As explained earlier, the edge contracts initially and bounces back from the $\rho=K_{\omega}$ toward the asymptotic region. Therefore, we found that the KK bubble cannot be formed from the gravitational collapse of matter we considered.

For generic cases, unlike above cases of $\omega=1/2$ and $2$, there is a naked singularity at $\rho=K_\omega$ since the light from the curvature singularity at $\rho=K_{\omega}$ can escape and be observed by an external observer within finite coordinate time. And the edge will approach to a naked singularity with no creation of an event horizon. However, since the future time-evolution around the singularity may involve strong quantum effects, 
the collapse process toward the spacetime with a naked singularity requires an introduction of other alternative theories, namely quantum gravity in order to treat this issue more properly.

In conclusion, it has been shown that the gravitationally collapsing clouds of matter successfully collapse to the GL black string solution while this is not the case for other class of solutions including the KK bubble solution and the spacetime with a naked singularity. The analysis has been achieved by an inhomogeneous matter distribution along the fifth direction, dealing with a careful consideration on the junction condition. There are still remained issues on the extraordinary geometrical properties of these solutions in the context of general relativity, which demands more studies on these geometries.
\vspace{1cm}

\textbf{Acknowledgments}\\
We are grateful to the referees of the journal for invaluable indications and suggestions 
on the issues of this paper. We would like to thank G Kang, S-H Yi, M S Yoon, and 
E J Son for helpful and fruitful discussions and comments. The work of S. Hyun was supported by the Basic Research Program of the Korea Science and Engineering Foundation under grant number R01-2004-000-10651-0 and by the Science Research Center Program of the Korea Science and Engineering Foundation through the Center for Quantum Spacetime (CQUeST) of Sogang University with grant number R11 - 2005 - 021. W Kim was supported by the Korea Research Foundation grant funded by the Korean Government (MOEHRD, Basic Research Promotion Fund) (KRF-2006-312-C00498). J Jeong was supported by the Korea Research Foundation Grant funded by Korea Government(MOEHRD, Basic Research Promotion Fund) (KRF-2005-070-C00030). J J Oh was supported by the Brain Korea 21(BK21) project funded by the Ministry of Education and Human Resources of Korea Government. 

\vspace{1cm}

\appendix
\begin{center}
{\bf \Large APPENDIX}
\end{center}
\section{Formation of the GL Black String in the Radial Coordinate System}\label{sec:appendix}
In this section, we study in more detail the formation of the GL black string from gravitationally collapsing dust cloud in the radial coordinate system which discussed in section \ref{sec:sols}. The crucial relations for the study are the matching conditions of the metric and the extrinsic curvatures between inside and outside the cloud, $[g_{ij}]=[K_{ij}]=0$. The interior metric ansatz is simply taken to be the metric (\ref{eq:inmet}) with $\xi=\xi(t)$ since the $zz$-components of the extrinsic curvatures in both sides vanish while the GL black string metric as an exterior metric is given in the form of
\begin{equation}
(ds)^2_{{\mathcal V}_{+}} = -\left(1-\frac{2M}{R}\right)dT^2 + \left(1-\frac{2M}{R}\right)^{-1}dR^2 + R^2 d\Omega^2 +dz^2.
\end{equation}
Since the analysis of the interior solution is similar to that in section \ref{sec:interior}, we shall investigate the behavior of the cloud edge from the junction conditions.
On the edge at $r=r_0$ and $R={\mathcal R}(t)$, the junction condition for metrics, $[g_{ij}]=0$, yields
\begin{equation}
\dot{T} = \frac{\mathcal R}{{\mathcal R}-2M}\sqrt{\dot{\mathcal R}^2 + 1-\frac{2M}{\mathcal R}}, ~~r_0^2 a^2 = {\mathcal R}^2.
\end{equation}

On the other hand, non-vanishing components of the extrinsic curvatures in the exterior coordinate system are found to be
\begin{equation}
K_{tt}^{+} = - \frac{d}{d\mathcal R} \sqrt{\dot{\mathcal R}^2 + 1-\frac{2M}{\mathcal R}},~~K_{\theta\theta}^{+}=K_{\phi\phi}^{+} =\frac{1}{\mathcal R} \sqrt{\dot{\mathcal R}^2 + 1-\frac{2M}{\mathcal R}},
\end{equation}
while they are evaluated to be
\begin{equation}
K_{\theta\theta}^{-}=K_{\phi\phi}^{-} = - \frac{1}{r_0 a},
\end{equation}
in the interior coordinate system. Therefore, the junction condition for the extrinsic curvature, $[K_{ij}]=0$, yields
\begin{equation}
\label{eq:twoeqs}
 \frac{d}{d\mathcal R} \sqrt{\dot{\mathcal R}^2 + 1-\frac{2M}{\mathcal R}} = 0, ~~\frac{1}{\mathcal R} \sqrt{\dot{\mathcal R}^2 + 1-\frac{2M}{\mathcal R}} + \frac{1}{r_0 a} = 0.
\end{equation}
The first equation in eq. (\ref{eq:twoeqs}) is simply solved to be
$\dot{\mathcal R}^2 + 1 - {2M}/{\mathcal R} = {\mathcal C}^2$ where ${\mathcal C}$ is an integration constant while the second equation is rewritten in the form of
$\dot{\mathcal R}^2 - {2M}/{\mathcal R} = 0$
by using the junction condition, $r_0^2a^2 = {\mathcal
  R}^2$. Comparing both equations yields ${\mathcal C}^2 =
1$ and one finally finds the equation of motion of the cloud edge,
\begin{equation}
\label{eq:eqsbs}
\dot{\mathcal R}^2 + V_{eff}(\mathcal R) = 0,
\end{equation}
where the effective potential is 
\begin{equation}
V_{eff}(\mathcal R) = - \frac{2M}{\mathcal R}.
\end{equation}
The effective potential is singular at ${\mathcal R}=0$ while it
vanishes at asymptotic region. Furthermore, it always has a negative
value in the whole region of ${\mathcal R}$.
The velocity of the edge at the horizon, ${\mathcal R}=2M$, does not
vanish while the acceleration of the edge is given in the form of
\begin{equation}
\ddot{\mathcal R} = - \frac{d}{2d{\mathcal R}} V_{eff}(\mathcal R) = - \frac{M}{{\mathcal R}^2}.
\end{equation}
Hence the edge will form an event horizon at ${\mathcal R}=2M$ and keep contracting until it forms a curvature singularity at ${\mathcal R}=0$.
An exact solution of eq. (\ref{eq:eqsbs}) is straightfowardly  obtained in the form of
\begin{equation}
{\mathcal R}(t) = \frac{1}{4M}\left[8({\mathcal R}_0 M)^{3/2} -
  12\sqrt{2}M^2 t\right]^{2/3},
\end{equation}
where ${\mathcal R}_0 \equiv {\mathcal R}(0)$.
The collapse time, $t_{c}$, at which the cloud edge approaches to a zero
size is found to be
\begin{equation}
t_{c} = \sqrt{\frac{2{\mathcal R}_0^3}{9M}}.
\end{equation}
which is clearly finite regardless of the initial data. 
Therefore, the edge of the collapsing cloud will form an event horizon at $t_h = t_c+4M/3$ and keep
contracting to form a curvature singularity at $R=0$ cloaked by an
event horizon. Provided $M=0$, describing the collapse in Minkowsi spacetime, then the
collapse time is no more finite, implying that the naked curvature
singularity at ${\mathcal R}=0$ never appear within finite comoving time.
\nc{\APP}[3]{Acta Phys. Polon. {\bf B#1}, (#3) #2}
\nc{\PR}[3]{Phys. Rev. {\bf #1}, (#3) #2}
\nc{\NPB}[3]{Nucl. Phys. {\bf B#1}, (#3) #2}
\nc{\PLB}[3]{Phys. Lett. {\bf B#1}, (#3) #2}
\nc{\PRD}[3]{Phys. Rev. {\bf D#1}, (#3) #2}
\nc{\PRL}[3]{Phys. Rev. Lett. {\bf #1}, (#3) #2}
\nc{\PREP}[3]{Phys. Rep. {\bf #1}, (#3) #2}
\nc{\EPJ}[3]{Eur. Phys. J. {\bf #1}, (#3) #2}
\nc{\PTP}[3]{Prog. Theor. Phys. {\bf #1}, (#3) #2}
\nc{\CMP}[3]{Comm. Math. Phys. {\bf #1}, (#3) #2}
\nc{\MPLA}[3]{Mod. Phys. Lett. {\bf A #1}, (#3) #2}
\nc{\CQG}[3]{Class. Quant. Grav. {\bf #1}, (#3) #2}
\nc{\NCB}[3]{Nuovo Cimento {\bf B#1}, (#3) #2}
\nc{\ANNP}[3]{Ann. Phys. (N.Y.) {\bf #1}, (#3) #2}
\nc{\GRG}[3]{Gen. Rel. Grav. {\bf #1}, (#3) #2}
\nc{\GC}[3]{Grav. Cosmol. {\bf #1}, (#3) #2}
\nc{\MNRAS}[3]{Mon. Not. Roy. Astron. Soc. {\bf #1}, (#3) #2}
\nc{\JHEP}[3]{JHEP {\bf #1}, (#3) #2}
\nc{\JCAP}[3]{JCAP {\bf #1}, (#3) #2}
\nc{\FP}[3]{Fortsch. Phys. {\bf #1}, (#3) #2}
\nc{\ATMP}[3]{Adv. Theor. Math. Phys. {\bf #1}, (#3) #2}
\nc{\AJP}[3]{Am. J. Phys. {\bf #1}, (#3) #2}
\nc{\ibid}[3]{{\it ibid.} {\bf #1}, (#3) #2}
\nc{\ZP}[3]{Z. Physik {\bf #1}, (#3) #2}
\nc{\PRSL}[3]{Proc. Roy. Soc. Lond. {\bf A#1}, (#3) #2}
\nc{\LMP}[3]{Lett. Math. Phys. {\bf #1}, (#3) #2}
\nc{\AM}[3]{Ann. Math. {\bf #1}, (#3) #2}
\nc{\hepth}[1]{{\rm [arXiv:hep-th/{#1}]}}
\nc{\grqc}[1]{{\rm [arXiv:gr-qc/{#1}]}}
\nc{\astro}[1]{{\rm [arXiv:astro-ph/{#1}]}}
\nc{\hepph}[1]{{\rm [arXiv:hep-ph/{#1}]}}
\nc{\phys}[1]{{\rm [arXiv:physics/{#1}]}}

\end{document}